\documentclass[acmsmall,screen]{acmart}
\usepackage{xspace}
\usepackage{tabularx}
\usepackage{tcolorbox}
\usepackage{xcolor}
\usepackage{adjustbox}
\usepackage[table]{xcolor}
\usepackage{pifont}
\usepackage{subcaption}
\usepackage{booktabs}
\usepackage{url}

\newcommand{\viberepair}{{\sc VibeRepair}\xspace}
\newcommand{\commentout}[1]{}
\newcommand{\circled}[1]{%
  \tikz[baseline=(char.base)]{
    \node[shape=circle,draw,inner sep=0.5pt,fill=black,text=white] (char) {#1};}}

\newcolumntype{Y}{>{\centering\arraybackslash}X}
\AtBeginDocument{%
  }

\begin{document}

\title{Specification Vibing for Automated Program Repair}

\author{Taohong Zhu}
\email{taohong.zhu@manchester.ac.uk}
\affiliation{%
  \institution{The University of Manchester}
  \city{Manchester}
  \country{United Kingdom}
}
\author{Lucas C. Cordeiro}
\email{lucas.cordeiro@manchester.ac.uk}
\affiliation{%
  \institution{The University of Manchester}
  \city{Manchester}
  \country{United Kingdom}
}
\author{Mustafa A. Mustafa}
\email{Mustafa.Mustafa@manchester.ac.uk}
\affiliation{%
  \institution{The University of Manchester}
  \city{Manchester}
  \country{United Kingdom}
}
\author{Youcheng Sun}
\email{youcheng.sun@mbzuai.ac.ae}
\affiliation{%
  \institution{Mohamed bin Zayed University of Artificial Intelligence}
  \city{Abu Dhabi}
  \country{UAE}
}

\begin{abstract}
  Large language model (LLM)-driven automated program repair (APR) has advanced rapidly, but most methods remain code-centric: they directly rewrite source code and thereby risk hallucinated, behaviorally inconsistent fixes. This limitation suggests the need for an alternative repair paradigm that relies on a representation more accessible to LLMs than raw code, enabling more accurate understanding, analysis, and alignment during repair. To address this gap, we propose \viberepair, a specification-centric APR technique that treats repair as behavior-specification repair rather than ad-hoc code editing. \viberepair first translates buggy code into a structured behavior specification that captures the program’s intended runtime behavior, then infers and repairs specification misalignments, and finally synthesizes code strictly guided by the corrected behavior specification. An on-demand reasoning component enriches hard cases with program analysis and historical bug–fix evidence while controlling cost. Across Defects4J and real-world benchmarks and multiple LLMs, \viberepair demonstrates consistently strong repair effectiveness with a significantly smaller patch space. On Defects4J v1.2, \viberepair correctly repairs 174 bugs, exceeding the strongest state-of-the-art baseline by 28 bugs, which corresponds to a 19\% improvement. On Defects4J v2.0, it repairs 178 bugs, outperforming prior approaches by 33 bugs, representing a 23\% improvement. Evaluations on real-world benchmarks collected after the training period of selected LLMs further confirm its effectiveness and generalizability. By centering repair on explicit behavioral intent, \viberepair reframes APR for the era of ``vibe'' coding: make the behavior sing, and the code will follow.
\end{abstract}

\ccsdesc[500]{Software and its engineering~Software maintenance tools}

\keywords{Automated Program Repair, Large Language Model, Behavior Specification, Prompt Engineering }

\maketitle

\section{INTRODUCTION}
Automated program repair (APR) aims to automatically generate patches for faulty programs, offering a promising means to reduce the substantial human effort required for debugging and bug fixing \cite{gazzola2018automatic,monperrus2018automatic}. By automating the repair process, APR has the potential to shorten development cycles and improve software reliability \cite{zhang2023survey}. As software defects remain unavoidable in practice, the ability to repair programs automatically has become an increasingly important research topic in software engineering \cite{monperrus2018living}.

Existing APR techniques can be broadly categorized into heuristic-based \cite{le2016history,le2011genprog,wen2018context}, constraint-based \cite{demarco2014automatic,le2017s3,long2015staged}, and template-based \cite{ghanbari2019practical,hua2018sketchfix,liu2019avatar} approaches, all of which aim to generate patches by directly modifying source code. While heuristic and constraint-based methods explore candidate fixes through search strategies or formal constraints, template-based techniques repair bugs using predefined fix patterns and can effectively handle many real-world defects. However, the reliance on manually designed templates and search heuristics limits their generalizability and scalability. To mitigate these issues, learning-based approaches—particularly those based on neural machine translation—have been widely studied, formulating program repair as a direct translation from buggy code to correct code using large bug-fixing datasets mined from open-source repositories~\cite{drain2021deepdebug,jiang2021cure,jiang2023knod,meng2023template,ye2022selfapr}. Despite their promise, these approaches remain inherently code-centric and strongly depend on data availability and quality, leaving fundamental challenges in generalization and semantic correctness unresolved \cite{zhang2023survey}.

Recent advances in large language models (LLMs) have substantially improved machines' ability to understand and generate source code, leading to growing interest in LLM-based APR. Existing LLM-based APR techniques typically generate candidate patches directly for faulty programs by leveraging LLMs' generative and reasoning capabilities. For example, ChatRepair \cite{xia2024automated} integrates patch generation with immediate feedback, enabling an interactive and conversational repair process. Thinkrepair \cite{yin2024thinkrepair} introduces chain-of-thought prompting ChatGPT to enhance defect reasoning during repair. ReinFix \cite{zhang2025repair} further guides LLM-based repair by extracting internal repair elements, such as key contextual information required to understand or fix a defect, as well as external repair elements derived from historically similar repair actions. These approaches demonstrate the effectiveness of enriching LLM-based repair with various forms of guidance.

Despite their differences, most existing LLM-based APR techniques share a common, code-centric repair paradigm: they analyze buggy programs and directly generate patched code. From the perspective of the software engineering lifecycle—spanning requirements, development, testing, and maintenance \cite{kute2014review} — many defects arise because the implemented behavior of a program deviates from the developer’s intended requirements \cite{ruan2024specrover}. Effective program repair, therefore, often requires capturing developer intent and using requirements to guide the repair process \cite{macaulay2012requirements,pohl1996requirements,zave1997four}. However, when LLMs are directly applied to transform buggy code into patched code, they may ignore, distort, or weaken critical requirement-related information embedded in the program~\cite{tian2025aligning}. This inaccurate perception of program intent fundamentally constrains the effectiveness of subsequent repair steps, as even refined repair strategies are built upon a flawed understanding of the underlying requirements. As a result, LLM-based program repair remains susceptible to hallucinated fixes, semantic inconsistencies, and overfitting, highlighting the limitations of purely code-centric repair approaches.

Inspired by the limitations of code-centric repair and the software engineering lifecycle, we propose \viberepair, a novel LLM-based APR approach that repairs programs by correcting behavior specifications. Rather than directly transforming buggy code into patched code, \viberepair enables LLMs to first translate a faulty program into a flawed behavior specification, infer the program’s intended behavior, realign and repair the flawed specification, and finally generate repaired code from the corrected specification.

To coordinate these steps and enable LLMs to accurately and autonomously complete this multi-stage repair process, we design a unified framework that structures program repair into three phases: transformation, repair, and generation. In the transformation phase, \viberepair employs carefully designed prompts and a behavior specification template to guide the LLM in converting a buggy program into a structured, flawed behavior specification. In the repair phase, a dedicated prompt guides the LLM to perform step-by-step analysis, including (1) inferring the correct intended behavior, (2) analyzing the root causes of specification misalignment and identifying how the specification should be realigned, and (3) producing a corrected behavior specification. In the generation phase, the LLM synthesizes repaired code from the corrected specification, ensuring that code is explicitly guided by repaired requirements rather than surface-level code patterns.

To further enhance repair effectiveness, \viberepair optionally incorporates a reasoning component during the repair phase. When enabled, this component provides additional reference information to assist the LLM in inferring intended behavior, identifying root causes, and formulating repair suggestions for specifications. Inspired by prior work on reasoning-guided repair, the reasoning component equips the LLM with a set of tools and access to a database of historical bug–fix pairs. Under a reasoning-and-acting framework, the LLM can autonomously plan and invoke tools as needed to obtain internal repair elements, such as analyzing function call traces during program execution, and external repair elements retrieved from the database based on defect similarity. These elements are supplied as auxiliary inputs to the repair phase, enabling more informed specification repair. Importantly, this component is activated only when the default repair path fails, allowing \viberepair to flexibly balance repair effectiveness, information relevance, and token cost while avoiding unnecessary distraction from critical repair signals.

In summary, this paper makes the following contributions: 
\begin{itemize}
    \item \textbf{Novel LLM-based Specification-Centric APR: }We propose a novel specification-centric approach to APR, which uniquely repairs behavior specifications before generating patch code, explicitly addressing specification misalignment rather than directly modifying buggy programs.

    \item \textbf{\viberepair Framework: }We design and implement \viberepair, an LLM-based APR framework that organizes the repair process into transformation, specification repair, and code generation phases, enabling LLMs to infer intended behavior, repair flawed behavior specifications, and synthesize patched programs accordingly.

    \item \textbf{Comprehensive Evaluation:} We conduct a comprehensive evaluation of \viberepair against state-of-the-art APR tools on the Defects4J \cite{just2014defects4j} and RWB \cite{yin2024thinkrepair} benchmarks. The results show that \viberepair consistently outperforms all evaluated baseline repair tools in terms of repair effectiveness. We further evaluate the generalizability of \viberepair across different LLMs, showing consistent improvements in repair effectiveness.

    \item \textbf{Open Science:} We open-source the complete \viberepair framework. Its specification-centric design is flexible and modular, allowing future work to readily incorporate new tools, analysis techniques, or reasoning strategies to explore improved APR solutions. Our source code is publicly available at \url{https://anonymous.4open.science/r/VibeRepair-9D73}.
\end{itemize}

\section{MOTIVATION}

\begin{figure}[h]
    \centering
    \includegraphics[width=\textwidth]{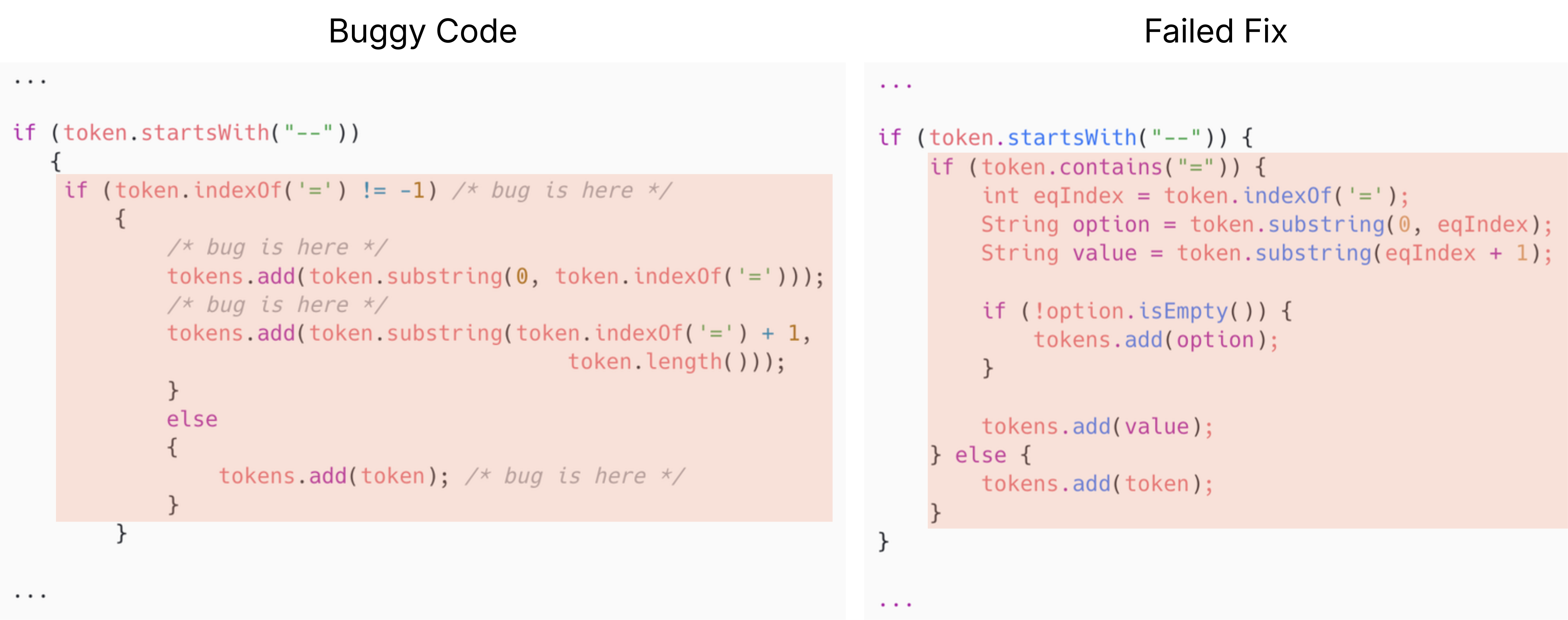}
    \caption{Buggy code and failed LLM-generated patch for {Cli-20} (Defects4J). \textbf{Left:} \texttt{flatten} unconditionally splits \texttt{-{}-x=y} at \texttt{`='} and appends both parts, even when \texttt{-{}-x} is not a recognized option, which can break \texttt{stopAtNonOption} handling for illegal options. \textbf{Right:} the LLM patch tweaks string handling but retains the same unconditional split, leaving the root cause unchanged.}
    \label{fig:fail_example}
\end{figure}

Figure~\ref{fig:fail_example} shows a buggy fragment from the \texttt{flatten} method in Cli-20 (Defects4J). The purpose of \texttt{flatten} is to normalize command-line arguments (e.g., \texttt{-{}-foo=bar}, \texttt{-abc}) into a canonical token sequence for downstream parsing, supporting both long and short options. When encountering an illegal or unrecognized option, the method is expected to respect the \texttt{stopAtNonOption} flag: if the flag is set, parsing should stop immediately and preserve the remaining arguments.

The Cli-20 bug arises because the implementation fails to check whether a long option is valid before splitting it. In particular, the buggy code always decomposes an argument such as \texttt{--foo=bar} into \texttt{-{}-foo} and \texttt{bar}, without first verifying that \texttt{-{}-foo} exists in the option set. As a result, unrecognized options are incorrectly decomposed and subsequently treated as if they were valid, which cascades into incorrect parsing behavior.

Figure~\ref{fig:fail_example} also shows an LLM-generated patch. Despite multiple repair attempts guided by validation-failure feedback, the patch does not address the root cause and instead makes a superficial null-handling change. This outcome highlights a key weakness of code-centric repair: without an explicit representation of the intended behavior of \texttt{flatten}, the model fails to recover the requirement being violated and therefore cannot reliably realign the implementation with that requirement.

\begin{figure}[h]
    \centering
    \includegraphics[width=\textwidth]{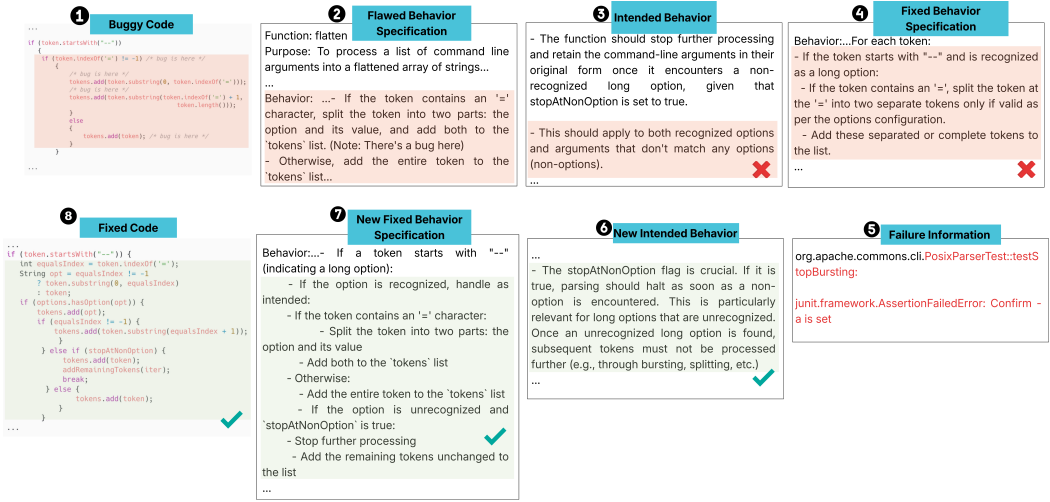}
    \caption{Specification-centric repair of Cli-20 using \viberepair. \viberepair translates the buggy \texttt{flatten} code ({1}) into an initial behavior specification ({2}), infers intended behavior ({3}), and produces a fixed specification used to generate a patch (rather than directly editing the code; {4}). When the patch fails validation, the failing test/error message ({5}) is used to refine the intended behavior ({6}) and revise the fixed specification ({7}). A second patch generated from the revised specification ({8}) passes all tests (red \ding{55} / green \checkmark).}
    \label{fig:sucess_example}
\end{figure}

Figure~\ref{fig:sucess_example} shows how \viberepair repairs Cli-20 by iteratively refining behavioral intent. Starting from the buggy code (\circled{1}), \viberepair generates an initial natural-language behavior specification that reflects the defect (\circled{2}). It then infers a candidate intended behavior (\circled{3}) and rewrites the defect specification accordingly to synthesize a patch (\circled{4}). This first patch fails validation (\circled{5}), indicating that the inferred intent is still incomplete. In particular, the model’s initial interpretation effectively treats an unrecognized long option as a signal to treat the remainder of the input as ordinary arguments, which can prematurely halt option parsing and mis-handle subsequent tokens (e.g., \texttt{-a}).

By incorporating this failure information, \viberepair enables the LLM to revise its understanding of the intended behavior. Specifically, the refined intent makes the role of \texttt{stopAtNonOption} explicit: options are processed left-to-right, and parsing halts immediately when a non-option is encountered under \texttt{stopAtNonOption}, without undoing or misclassifying options already parsed (\circled{6}). This refined intent is encoded into an updated fixed specification (\circled{7}), which guides a second round of code generation. The resulting patch (\circled{8}) passes all validation tests.

Overall, this example shows that successful repair can depend on recovering the correct behavioral intent. By externalizing intent as an explicit behavior specification and updating it using validation feedback, \viberepair corrects the specification first, allowing the implementation to follow.

\section{VIBEREPAIR}

\begin{figure*}[h]
    \centering
    \includegraphics[width=\textwidth]{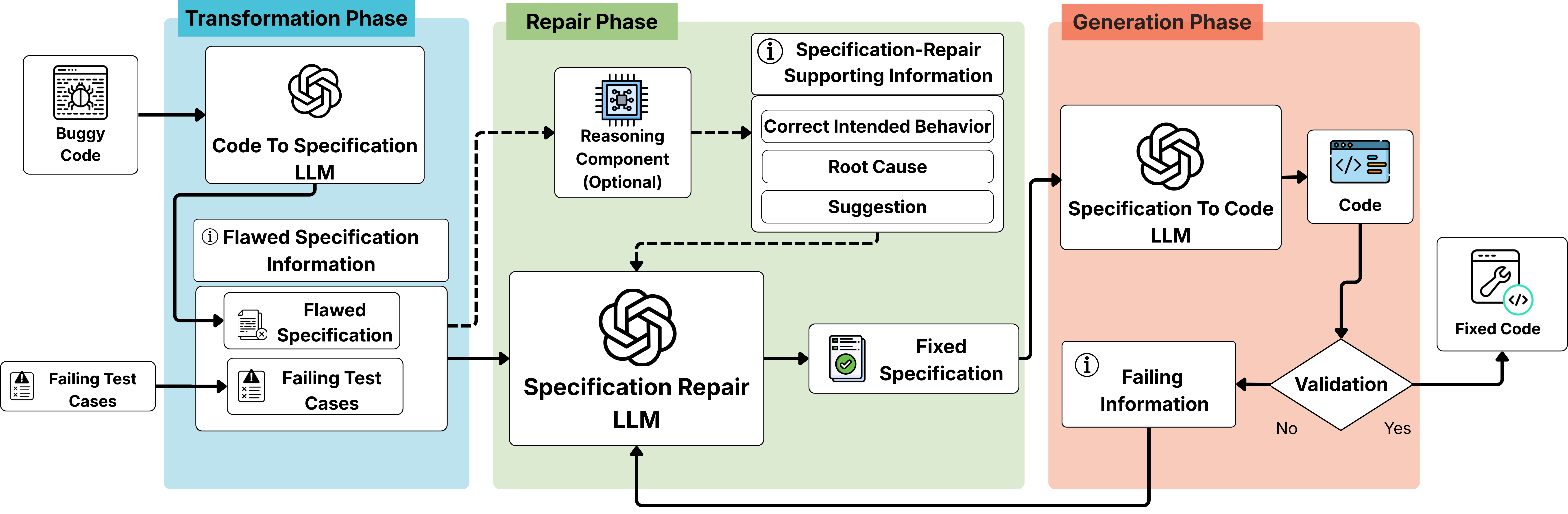}
    \caption{Overview of \viberepair. \viberepair proceeds in three phases: 1) Transformation: an LLM translates buggy code into a structured, flawed behavior specification that follows a predefined template and combines it with failing tests to form the repair input. 2) Repair: the LLM corrects specification misalignments to produce a fixed specification, optionally using an on-demand reasoning component that provides additional repair supporting information. 3) Generation: the LLM synthesises candidate code from the fixed specification and validates it against the test suite; when validation fails, failing test information is fed back to the repair phase to refine the specification and repeat repair and generation until all tests pass.}
    \label{fig:overview}
\end{figure*}

In this section, we present \viberepair, an LLM-based APR approach designed to improve the repair effectiveness through a specification-centric design. As illustrated in Fig.~\ref{fig:overview}, \viberepair organises the repair process into three phases: the transformation phase, the repair phase, and the generation phase.

\commentout{
In the transformation phase, \viberepair employs an LLM to translate a buggy program into a flawed behavior specification that conforms to a predefined specification template. This flawed specification, together with the failing test cases that the buggy program does not satisfy, forms the flawed specification information, which serves as the primary input for subsequent repair. In the repair phase, \viberepair repairs the flawed specification to produce a fixed specification. By default, the LLM directly repairs the flawed specification using the information from that specification. Optionally, a reasoning component can be enabled to generate additional specification-repair supporting information, which is then incorporated to guide the LLM’s analysis and improve specification repair. In the generation phase, the LLM generates candidate code from the fixed specification and validates it against the test suite. If all test cases pass, the generated code is returned as the final patch. Otherwise, the failing test information is fed back to the repair phase, where the LLM performs further analysis, updates the specification accordingly, and iteratively repeats the repair and generation process until a correct patch is produced.
}

\subsection{Transformation Phase}
\begin{figure}[h]
    \centering
    \includegraphics[width=\textwidth]{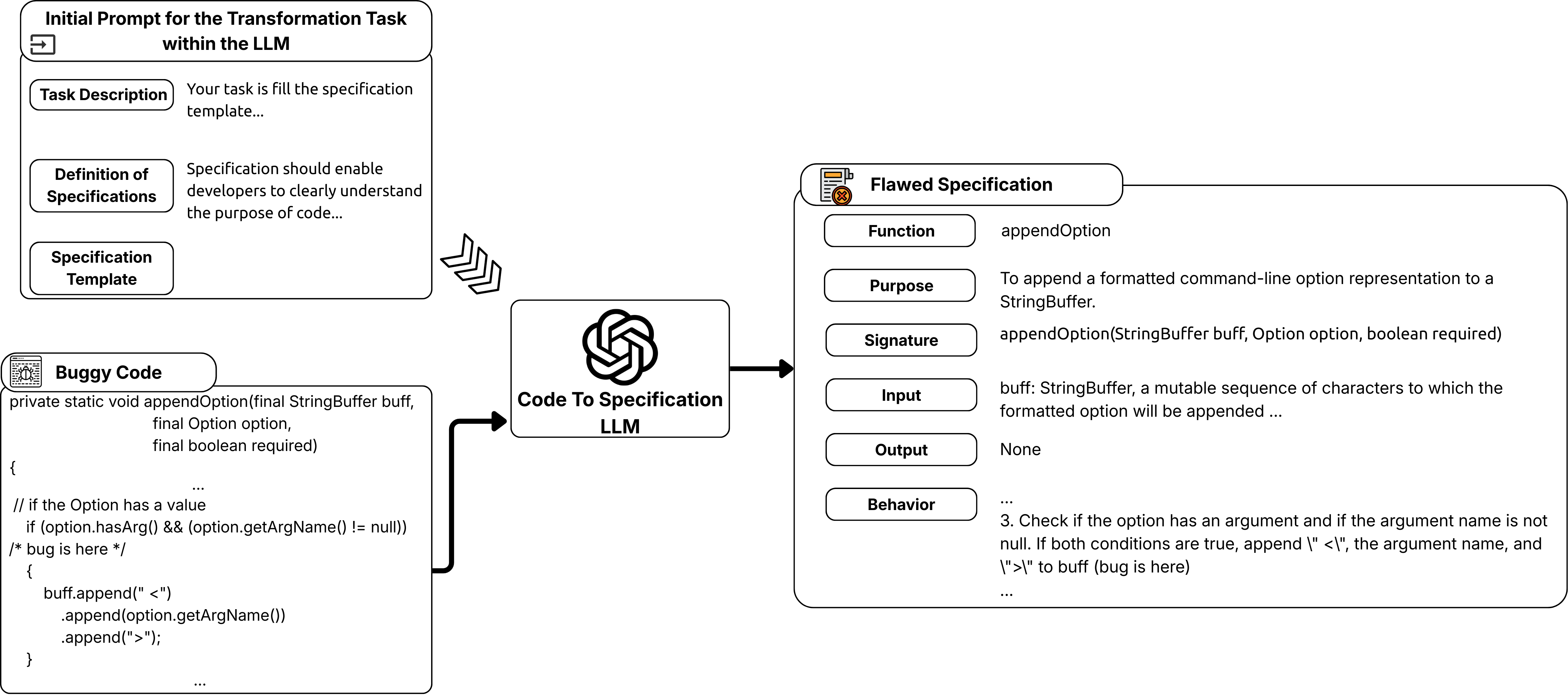}
    \caption{Transformation phase: given the buggy code, the model fills a specification template to produce an initial natural-language behavior specification that inherits the defect.}
    \label{fig:tran_phase}
\end{figure}

The goal of the transformation phase is to enable the LLM to translate buggy code into a structured behavior specification that captures the intended behavior reflected by the buggy code. To this end, we design a zero-shot \cite{kojima2022large} initial prompt that is internally set within the LLM to guide the code-to-specification transformation, without requiring any task-specific fine-tuning. As illustrated in Fig.~\ref{fig:tran_phase}, the prompt consists of three main components: Task Description, Definition of Specifications, Specification Template.

The first component is a task description that instructs the LLM to identify the intent of the given code and complete a behavior specification using a predefined template and its corresponding guidelines. The second component provides a high-level definition of behavior specifications, emphasizing that a completed specification should clearly describe the code's purpose and enable reimplementation with the same intended behavior.

The third component is the specification template, which defines the structure of the output specification. We design the following specification template:
\begin{tcolorbox}[
  colback=gray!10,
  colframe=black!30,
  boxrule=0.5pt,
  arc=2mm,
  left=6pt,
  right=6pt,
  top=6pt,
  bottom=6pt
]
\textcolor{red}{\textbf{Function: }}
\#Function name

\medskip
\textcolor{red}{\textbf{Purpose: }}
\#Purpose of this function

\medskip
\textcolor{red}{\textbf{Signature: }}
\#The unique identifier of a function, defined by its name and parameter list (and sometimes its return type, depending on the language)

\medskip
\textcolor{red}{\textbf{Input: }}
\#Input parameters of the function, including their types, and explain the meaning of "input".

\medskip
\textcolor{red}{\textbf{Output: }}
\#Output parameters of the function, including their types, and explain the meaning of "output".

\medskip
\textcolor{red}{\textbf{Behavior: }}
\#Explain step by step the behavior of a function during its execution. If a comment in the code indicates that a certain part has a bug, after the corresponding behavior, add a parenthesis containing additional information indicating that this behavior has a bug.

\end{tcolorbox}
Each attribute in the template is accompanied by detailed explanations and guidelines specifying what information should be filled in, guiding the LLM to systematically extract and organize relevant information from the buggy code into a well-structured specification.

With this prompt design, \viberepair only requires the buggy program as input. The LLM automatically produces a flawed behavior specification. This flawed specification is then combined with the failing test cases of the buggy program to form the flawed specification information, which serves as the input to the subsequent repair phase.

\subsection{Repair Phase}

\subsubsection{Default Specification Repair}
\label{subsec: Specification Repair}
\begin{figure}[h]
    \centering
    \includegraphics[width=\textwidth]{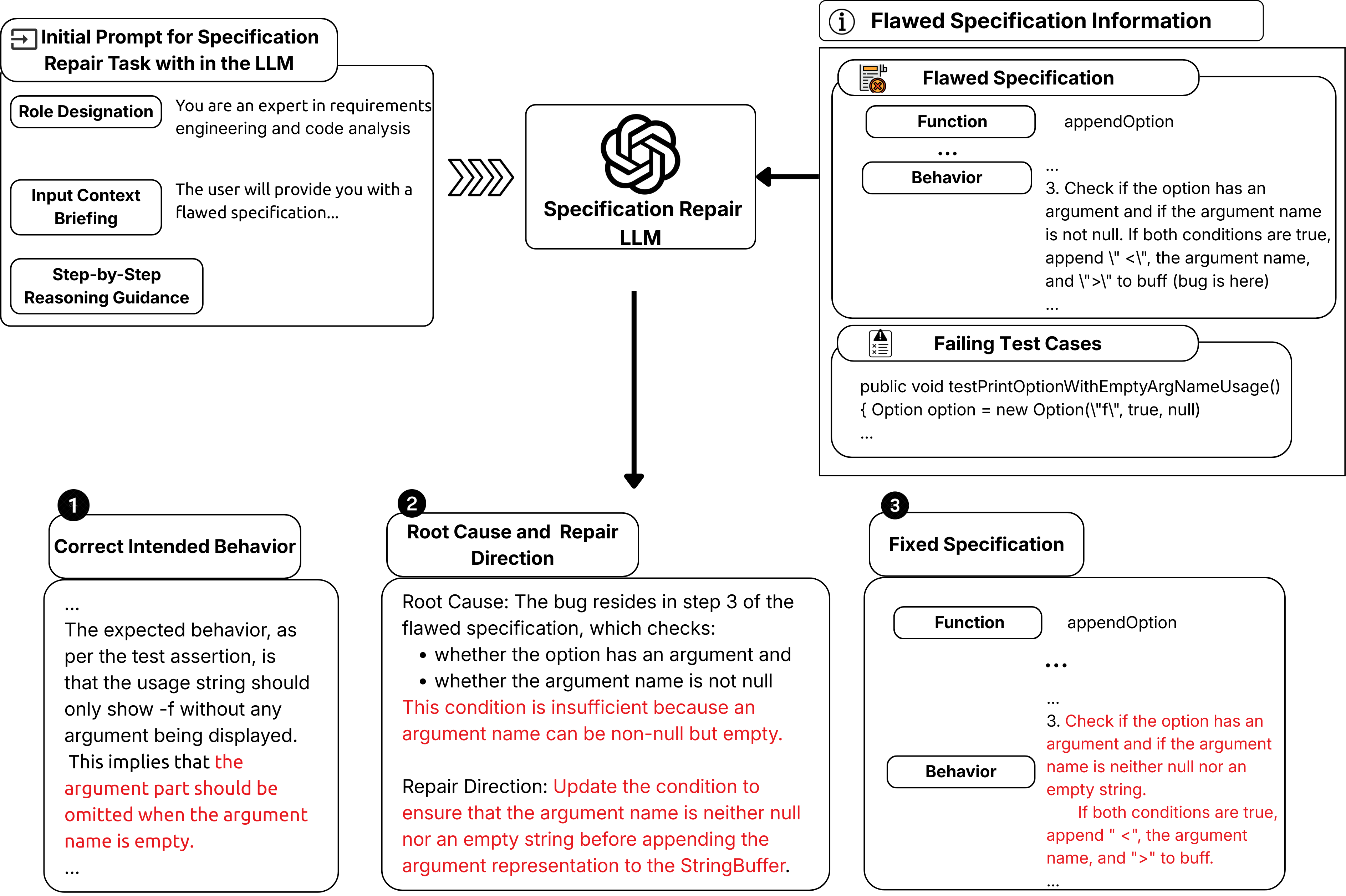}
    \caption{Fixing a flawed specification in the repair phase. Given the flawed specification information, which consists of a flawed specification and the corresponding failing test cases, the model infers the intended behavior, identifies the root cause and repair direction, and outputs a revised specification.}
    \label{fig:repair_phase}
\end{figure}

In the repair phase, the flawed specification information produced in the transformation phase is provided to an LLM to perform the specification repair task. To reliably steer the LLM toward structured, reproducible repairs, we design a dedicated initial prompt by combining a zero-shot prompting setup with step-by-step reasoning guidance, following the chain-of-thought prompting \cite{wei2022chain} methodology. As shown in Fig.~\ref{fig:repair_phase}, the prompt consists of three main components: Role Designation, Input Context Briefing, and Step-by-Step Reasoning Guidance.

\textbf{Role Designation} assigns the LLM an explicit expert role (e.g., “You are an expert in requirements engineering and code analysis”), encouraging it to focus on requirement-level reasoning rather than ad-hoc code edits.

\textbf{Input Context Briefing} explains the artifacts the LLM will receive and how they relate to the repair task. Specifically, the prompt states that (i) the flawed specification describes the intended behavior of a target function and is used to guide development, (ii) the implementation developed from this flawed specification fails certain test cases, and (iii) the failing test cases, together with their corresponding error messages, are provided as additional evidence of the behavioral mismatch.

\textbf{Step-by-Step Reasoning Guidance} structures the repair process into three steps:
\begin{enumerate}
    \item Infer the correct intended behavior. The LLM first infers the target function's intended behavior by analyzing failing test cases, identifying the behavioral expectations that are violated.

    \item Diagnose the root cause and derive repair directions. Given the inferred intended behavior, the LLM then reasons about where and why the flawed specification deviates from the intended behavior, diagnoses the root cause of the misalignment, and derives concrete directions for realigning the specification.

    \item Produce the repaired specification. Finally, the LLM synthesizes a fixed specification by applying the derived repair directions, producing a corrected behavior specification that is aligned with the intended behavior.
\end{enumerate}

With this prompt design, once the repair-phase LLM receives the flawed specification information, it can automatically conduct intent inference, root-cause analysis, and specification revision to generate a fixed specification, which is then passed to the subsequent generation phase for code synthesis and validation.

\subsubsection{Optional Reasoning Component}
While the default repair process is sufficient for many defects, \viberepair additionally provides an optional Reasoning Component that is activated when the default repair process fails to successfully repair the specification. This component supplies additional evidence to assist specification repair in challenging cases. We intentionally exclude it from the default workflow for two reasons. First, executing the reasoning process involves multiple tool invocations and interactions with external resources, which can significantly increase token consumption and repair cost. Second, introducing excessive auxiliary information may distract the repair LLM from critical signals, potentially shifting its focus and degrading repair accuracy. By enabling this component only upon repair failure, \viberepair balances repair effectiveness, efficiency, and information relevance.

\begin{figure}[h]
    \centering
    \includegraphics[width=0.9\textwidth]{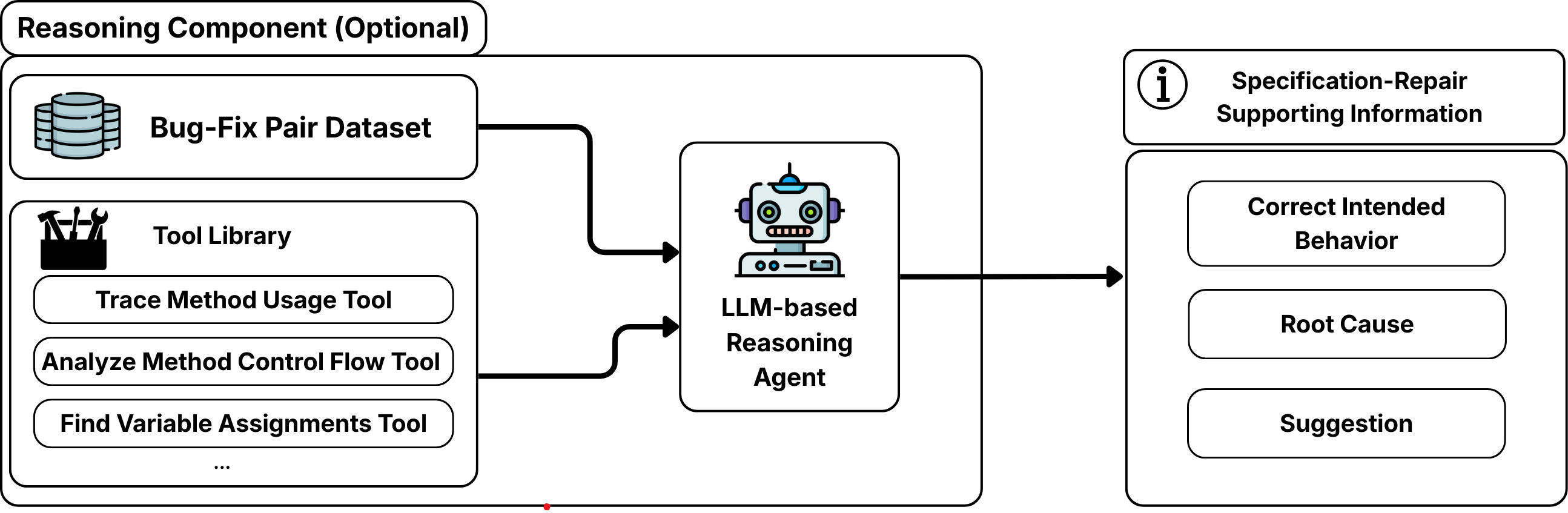}
    \caption{Reasoning component in \viberepair. An LLM-based reasoning agent consults a tool library and bug--fix pairs to generate supporting information for specification repair (intended behavior, root cause, and suggestions).}
    \label{fig:Reasoning_Component}
\end{figure}

Figure \ref{fig:Reasoning_Component} illustrates the internal architecture of the Reasoning Component. Motivated by the reasoning-agent design proposed by Zhang et al. and leveraging their open-source resources \cite{zhang2025repair}, we implement an LLM-based reasoning agent under a Reasoning-and-Acting framework \cite{yao2022react}. The agent is equipped with a tools library (Table 1) and a large example database containing tuples of ($buggy\_code$, $fix\_code$, $root\_cause$, $embedding\_value$). The $embedding\_value$ is computed by concatenating $buggy\_code$ and $root\_cause$ and transforming the combined text into an embedding vector. To ensure the agent performs reasoning in a controlled and reproducible manner, we design a dedicated agent prompt that specifies the agent's role, the available tools and their usage constraints, and a step-by-step action plan. Driven by this prompt, the agent iteratively alternates between reasoning and acting, invoking tools only when needed and following the prescribed steps to produce structured outputs.

\begin{table*}[t]
\caption{Overview of Tools in the Tools Library \cite{zhang2025repair}.}
\centering
\footnotesize
\renewcommand{\arraystretch}{1.05}
\captionsetup{skip=6pt}
\begin{tabularx}{\textwidth}{p{0.21\textwidth} X}
\toprule
\textbf{Tool Name} & \textbf{Description} \\
\midrule
example\_retrieval &
Retrieves repair examples that address root causes similar to flawed specifications \\
identify\_variable &
Examines how variables are declared and referenced within designated files \\
find\_variable\_assignments &
Identifies all locations where a given variable receives a value \\
track\_variable\_dataflow &
Analyzes the propagation of variable values from their origins to their destinations \\
trace\_method\_usage &
Determines where and in which files a specified method is invoked \\
analyze\_method\_details &
Provides an in-depth analysis of a method’s structure and characteristics \\
find\_method\_in\_file &
Examines the internal control-flow constructs of a method, such as conditionals and loops \\
find\_class\_loc &
Determines the source location in which a class is defined \\
identify\_class &
Extracts the complete class definition, including its fields and methods \\
get\_imports &
Collects all import declarations present in a specified file \\
\bottomrule
\end{tabularx}
\label{tab:tools}
\end{table*}

Given the flawed specification information, the agent executes the prompt-defined procedure as follows:
\begin{enumerate}
    \item It invokes tools to query project-level information (e.g., function invocation contexts and execution traces) to infer the correct intended behavior of the target function described by the flawed specification.

    \item Conditioned on the inferred intended behavior, it further analyzes the project context to diagnose the root cause of the specification misalignment that leads to the observed test failures.

    \item It invokes an $example\_retrieval$ tool to search for relevant external repair evidence. Specifically, the tool converts the target ($buggy\_code$, $root\_cause$) into an embedding vector q and computes its similarity with the stored $embedding\_value$ entries in the database. If the highest similarity exceeds a threshold of 0.6, the corresponding ($buggy\_code$, $fix\_code$, $root\_cause$) tuple is retrieved as an example; otherwise, no example is returned.

    \item When an example is available, the agent uses it to cross-check and refine the inferred intended behavior and root cause, and then derives a concrete specification repair suggestion describing how the flawed specification should be revised.
\end{enumerate}

The reasoning outcomes—including the inferred correct intended behavior, diagnosed root cause, and generated repair suggestion—are collected and organized as Specification-Repair Supporting Information. This artifact is returned by the Reasoning Component and provided to the repair-phase LLM as additional guidance for repairing the flawed specification. Through this on-demand, prompt-driven, and evidence-informed reasoning process, \viberepair strengthens specification repair in difficult cases while avoiding unnecessary cost and information overload when the default repair process is sufficient.

\subsection{Generation Phase}
In the generation phase, the fixed specification produced by the repair phase is provided to an LLM to synthesize executable code. To guide this process, we include a clear task description in the prompt, instructing the LLM to generate code that strictly adheres to the fixed specification. This prompt design enables the LLM to automatically generate code from the fixed specification upon receipt, ensuring that code synthesis is explicitly guided by the repaired specification.

As illustrated in the Generation Phase of Fig.~\ref{fig:overview}, after code generation, the synthesized code is validated against the existing test suite. If the generated code passes all test cases, it is returned as a patch for the buggy program. This patch constitutes both the output of the generation phase and the final output of \viberepair.

If the generated code fails validation, the failure information is fed back to the repair phase to trigger another round of specification repair. The failure information includes the test cases that fail, the expected and actual outputs observed during test execution, and, when applicable, compilation error messages if the code does not successfully compile. Upon receiving this information, the Specification-Repair LLM (Section \ref{subsec: Specification Repair}) revisits the specification repair task, leveraging its prior interaction history and the newly observed failure evidence under the same step-by-step reasoning guidance. The LLM then revises the specification accordingly to address the uncovered inconsistencies.

This iterative feedback loop between the repair and generation phases continues until a valid patch is produced or a predefined termination condition is met. Through this design, \viberepair incrementally refines behavior specifications based on concrete execution feedback, allowing specification repair and code generation to jointly converge toward a correct patch.

\section{EVALUATION DESIGN}
\subsection{Research Questions}
To evaluate the effectiveness and generalizability of \viberepair, we formulate three research questions. Together, they examine (i) overall repair effectiveness across datasets and repair settings, (ii) the effect of enabling the optional reasoning component under different usage strategies and cost considerations, and (iii) generalization across different LLMs.

\begin{itemize}
    \item \textbf{RQ1: How effective is \viberepair across different datasets and repair scenarios compared to existing APR techniques?} This research question evaluates the overall repair effectiveness of \viberepair by comparing it against multiple state-of-the-art APR baseline tools under diverse datasets and repair scenarios. The goal is to assess whether \viberepair consistently outperforms existing approaches across different benchmarks, providing a comprehensive view of its repair capability.

    \item \textbf{RQ2: What is the impact of different component usage strategies on repair effectiveness and cost?} This research question examines how enabling the optional reasoning component under different strategies affects the repair effectiveness of \viberepair and associated cost. The goal is to understand the trade-offs between repair performance and computational overhead, and to evaluate whether selectively activating the reasoning component can improve repair outcomes while controlling costs.

    \item \textbf{RQ3: How well does \viberepair generalize across different LLMs?} This research question investigates the generalizability of \viberepair by evaluating its repair effectiveness when using different LLMs. The goal is to determine whether the proposed approach is robust to the choice of underlying LLM and can consistently improve repair performance across models with varying capabilities.

\end{itemize}

\subsection{Benchmarks}
To evaluate the effectiveness of \viberepair, we conduct experiments on multiple widely used datasets that serve as benchmarks in APR research.

\textbf{Defects4J.} To ensure accurate, effective, and fair comparisons with existing APR techniques, we adopt Defects4J, a widely used benchmark dataset in the APR community. Following prior studies, we evaluate \viberepair under the same experimental settings to facilitate direct comparison with previously reported results. In particular, we divide Defects4J into two versions, Defects4J v1.2 and Defects4J v2.0. Defects4J v1.2 consists of 391 bugs collected from six Java projects, while Defects4J v2.0 extends the benchmark with 438 additional bugs from nine more projects. Following established evaluation practices, we further categorize Defects4J bugs into four repair scenarios: multi-function (MF), single-function (SF), single-hunk (SH), and single-line (SL). The distribution of bugs across different repair scenarios in Defects4J v1.2 and V2.0 is summarized in Table \ref{tab:Statistics_Defects4J}. When evaluating the effectiveness of \viberepair's repairs, we include bugs across all repair scenarios to provide a comprehensive assessment of its repair capability.

\begin{table*}[h]
\caption{Statistics of Defects4J}
\centering
\footnotesize
\renewcommand{\arraystretch}{1.1}
\captionsetup{skip=6pt}
\begin{tabularx}{0.82\textwidth}{p{0.12\textwidth} *{5}{Y}}
\toprule
\textbf{Dataset} &
\textbf{\# Total Bugs} &
\textbf{\# MF Bugs} &
\textbf{\# SF Bugs} &
\textbf{\# SH Bugs} &
\textbf{\# SL Bugs} \\
\midrule
Defects4J v1.2 & 391 & 136 & 255 & 154 & 80 \\
Defects4J v2.0 & 438 & 210 & 228 & 159 & 78 \\
\bottomrule
\end{tabularx}
\label{tab:Statistics_Defects4J}
\end{table*}

\textbf{RWB.} To mitigate the risk of data leakage and to evaluate the generalizability of \viberepair under stricter conditions, we additionally adopt two Real-World Bugs (RWB) datasets collected by Yin et al. \cite{yin2024thinkrepair}, which have been used in prior studies. According to the latest released dataset statistics, RWB v1.0 contains 27 bugs collected from five Java projects, mined from bug-fixing commits after October 2021. RWB v2.0 also includes 27 bugs drawn from the same set of Java projects as v1.0, collected from commits after March 2023.

\subsection{Baselines and Evaluation Metrics}
\textbf{Baselines.} We compare \viberepair with a set of representative APR techniques, including both LLM-based and traditional approaches. The LLM-based baselines include ReinFix \cite{zhang2025repair}, ChatRepair \cite{xia2024automated}, ThinkRepair \cite{yin2024thinkrepair}, RepairAgent \cite{bouzenia2024repairagent}, FitRepair \cite{xia2023revisiting}, AlphaRepair \cite{xia2022less}, and RAP-Gen \cite{wang2023rap}, while the traditional baselines include GAMMA \cite{zhang2023gamma}, TENURE \cite{meng2023template}, Tare \cite{zhu2023tare}, KNOD \cite{jiang2023knod}, and Recoder \cite{zhu2021syntax}. Following common practice in APR evaluations \cite{zhang2025repair,xia2024automated,yin2024thinkrepair,bouzenia2024repairagent,xia2023revisiting,xia2022less,wang2023rap,zhang2023gamma,meng2023template,zhu2023tare,jiang2023knod,zhu2021syntax}, we reuse the repair results reported in prior studies for these baseline techniques rather than re-running all tools.

\textbf{Evaluation Metrics.} We evaluate repair effectiveness using two standard metrics: the number of plausible patches and the number of correct patches. A plausible patch is one that passes the entire test suite, and a correct patch is a plausible patch that is semantically or syntactically equivalent to the reference developer patch, as determined through manual inspection. LLM usage cost is estimated from token consumption using public pricing rates, with GPT-4o priced at \$5 per million input tokens and \$15 per million output tokens, and GPT-4 priced at \$30 per million input tokens and \$60 per million output tokens.

\subsection{Implementation}

\viberepair is implemented on top of the LangChain framework \cite{langchain2024}. For evaluation on Defects4J, we use GPT-4o \cite{gpt4o} as the base model. When evaluating on RWB v1.0, we use GPT-3.5 \cite{gpt3.5} and GPT-4 \cite{gpt4}, and for RWB v2.0, we adopt DeepSeek-Coder \cite{deepseek} as the base model. For all LLMs, the temperature is set to 1. Following prior studies \cite{zhang2025repair,yin2024thinkrepair,xia2024automated}, we eliminate bias introduced by different fault localization techniques by conducting all experiments under perfect fault localization.

During patch generation, we limit the number of repair attempts to at most five per bug. For each attempt, the repair-phase LLM can receive up to three rounds of feedback from the generation phase when validation fails. As a result, the total maximum patch space size is 5 × 3 = 15.

\section{EVALUATION RESULTS}

\subsection{RQ1 - Repair Effectiveness}
To answer RQ1, we evaluate the repair effectiveness of \viberepair on the Defects4J v1.2 and Defects4J v2.0 benchmarks. We compare \viberepair against a set of state-of-the-art APR tools to provide a comprehensive assessment by directly comparing it with existing approaches. To reduce potential bias introduced by differences in LLM capabilities, we configure \viberepair to use the same base model as ReinFix, namely GPT-4o, in all experiments for this research question. In the following evaluation, \viberepair\textsubscript{\scriptsize miniR\_GPT4o} denotes the configuration where \viberepair uses GPT-4o as the base model and adopts a minimal reasoning strategy, in which the optional Reasoning Component is activated only when the default repair process fails to repair a bug. We refer to this configuration as \viberepair\textsubscript{\scriptsize miniR\_GPT4o} throughout the remainder of this section. In contrast, \viberepair\textsubscript{\scriptsize GPT4o} denotes the configuration that uses GPT-4o as the base model but does not enable the Reasoning Component. We refer to this configuration as \viberepair\textsubscript{\scriptsize GPT4o} in the following discussion.

\begin{table*}[t]
\caption{Overall repair performance of \viberepair and baseline APR tools on Defects4J, reported per project and in total. Each entry is shown as \#correct/\#plausible fixes, and the header reports each tool’s sampling budget (the number of candidate patches generated per bug), ``--'' indicates that plausible-fix counts were not reported for that tool.} 
\centering
\footnotesize
\renewcommand{\arraystretch}{1.25}
\setlength{\extrarowheight}{2pt}
\setlength{\tabcolsep}{2pt}
\captionsetup{skip=6pt}

\begin{adjustbox}{max width=\textwidth}
\begin{tabular}{p{2.2cm} *{14}{c}}
\toprule
\textbf{APR Tool} &
\rotatebox{85}{\textbf{\viberepair}\textsubscript{\scriptsize miniR\_GPT4o}} &
\rotatebox{85}{\textbf{\viberepair}\textsubscript{\scriptsize GPT4o}} &
\rotatebox{85}{\textbf{ReinFix}\textsubscript{\scriptsize GPT4o}} &
\rotatebox{85}{\textbf{ChatRepair}} &
\rotatebox{85}{\textbf{ThinkRepair}} &
\rotatebox{85}{\textbf{RepairAgent}} &
\rotatebox{85}{\textbf{FitRepair}}  &
\rotatebox{85}{\textbf{GAMMA}} &
\rotatebox{85}{\textbf{TENURE}} &
\rotatebox{85}{\textbf{Tare}} &
\rotatebox{85}{\textbf{AlphaRepair}} &
\rotatebox{85}{\textbf{RAP-Gen}} &
\rotatebox{85}{\textbf{KNOD}} &
\rotatebox{85}{\textbf{Recoder}} \\
\midrule

\textbf{Sampling Times} &
\cellcolor{gray!20} \textbf{5$\times$3} & \textbf{5$\times$3} & 3$\times$3$\times$5 & 500 & 25$\times$5 & 117 &
1000$\times$4 & 250 & 500 & 100 & 5000 & 100 & 1000 & 100 \\
\midrule

Chart   & \cellcolor{gray!20}20/21 & 20/21 & 18/20 & 15/- & 11/- & 11/14 & 8/-  & 11/11 & 7/-  & 11/- & 9/-  & 9/-  & 10/11 & 9/- \\
Closure & \cellcolor{gray!20}44/54 & 35/42 & 40/50 & 37/- & 31/- & 25/25 & 29/- & 24/26 & 26/- & 25/- & 23/- & 22/- & 23/29 & 25/- \\
Lang    & \cellcolor{gray!20}40/49 & 31/37 & 33/47 & 21/- & 19/- & 17/17 & 19/- & 16/25 & 16/- & 14/- & 13/- & 12/- & 11/13 & 12/- \\
Math    & \cellcolor{gray!20}51/76 & 45/66 & 39/68 & 32/- & 27/- & 29/29 & 24/- & 25/31 & 22/- & 22/- & 21/- & 26/- & 20/25 & 20/- \\
Mockito & \cellcolor{gray!20}9/11 & 6/6 & 10/11 & 6/-  & 6/-  & 6/6   & 6/- & 3/3   & 4/-  & 2/-  & 5/-  & 2/-  & 5/5   & 2/-  \\
Time    & \cellcolor{gray!20}10/12 & 10/11 & 6/11  & 3/-  & 4/-  & 2/3   & 3/-  & 3/5   & 4/-  & 3/-  & 3/-  & 1/-  & 2/2   & 3/- \\
\midrule

\textbf{\#Total (D4J v1.2)} &
\cellcolor{gray!20}\textbf{174}/\textbf{223} & \textbf{147}/183 & 146/207 & 114/- & 98/- & 90/94 & 89/- & 82/101 & 79/- & 77/- & 74/109 & 72/- & 71/85 & 71/- \\
\textbf{\#Total (D4J v2.0)} &
\cellcolor{gray!20}\textbf{178/223} & \textbf{167/208} & 145/190 & 48/- & 107/- & 74/92 & 44/- & 45/- & 50/- & -/- & 36/- & 53/- & 50/85 & 11/- \\
\bottomrule

\end{tabular}
\end{adjustbox}

\label{tab:Defects4J_result}
\end{table*}

\subsubsection{Overall performance.} For Defects4J v1.2, Table \ref{tab:Defects4J_result} shows that \viberepair\textsubscript{\scriptsize miniR\_GPT4o} achieves the strongest overall performance, generating 174 correct patches out of 223 plausible patches. Compared to ReinFix\textsubscript{\scriptsize GPT4o}, which produces 146 correct patches out of 207 plausible patches, \viberepair\textsubscript{\scriptsize miniR\_GPT4o} repairs 28 more bugs correctly and produces 16 additional plausible patches. Even without enabling the Reasoning Component, \viberepair\textsubscript{\scriptsize GPT4o} remains highly competitive, yielding 147 correct patches out of 183 plausible patches. Although this configuration produces fewer plausible patches than ReinFix, it still repairs one more bug correctly, indicating that a larger proportion of its generated patches are correct. This suggests that \viberepair\textsubscript{\scriptsize GPT4o} tends to generate a smaller set of candidate patches with higher repair precision, whereas ReinFix produces more plausible patches overall but with a higher rate of incorrect repairs. At the project level, \viberepair\textsubscript{\scriptsize miniR\_GPT4o} shows particularly strong improvements over ReinFix on projects such as Math, where it repairs 51 bugs compared to the 39 of ReinFix, and Time, where it repairs 10 bugs compared to the 6 of ReinFix.

On Defects4J v2.0, the advantage of \viberepair becomes even more pronounced. As shown in Table \ref{tab:Defects4J_result}, \viberepair\textsubscript{\scriptsize miniR\_GPT4o} repairs 178 bugs correctly, whereas ReinFix\textsubscript{\scriptsize GPT4o} repairs 145 bugs, resulting in 33 additional correct repairs. In terms of plausible patches, \viberepair\textsubscript{\scriptsize miniR\_GPT4o} produces 223 plausible patches, which is 33 more than the 190 plausible patches produced by ReinFix. \viberepair\textsubscript{\scriptsize GPT4o} also substantially outperforms ReinFix, repairing 167 bugs correctly and producing 208 plausible patches, corresponding to 22 more correct repairs and 18 more plausible patches, respectively. These results indicate that the proposed specification-centric repair pipeline is effective even without the optional Reasoning Component. Importantly, all these improvements are achieved with the smallest exploration budget among the compared approaches: \viberepair explores a maximum patch space size of 5 × 3, whereas existing baselines rely on significantly larger sampling budgets, often ranging from hundreds to thousands of candidate patches.

\begin{table*}[!htp]
\caption{Repair results (correct fixes) on Defects4J v1.2 and v2.0 under four repair scenarios. MF and SF group bugs by the scope of the fix (multi-function vs single-function), while SH and SL group bugs by patch granularity (single-hunk vs single-line).}
\centering
\footnotesize
\renewcommand{\arraystretch}{1.3}
\setlength{\extrarowheight}{2pt}
\setlength{\tabcolsep}{4.5pt}  
\captionsetup{skip=6pt}

\begin{adjustbox}{max width=\textwidth}
\begin{tabular}{p{2.8cm} cccc cccc} 
\toprule
\textbf{Benchmarks} &
\multicolumn{4}{c}{\textbf{Defects4J v1.2}} &
\multicolumn{4}{c}{\textbf{Defects4J v2.0}} \\
\cmidrule(lr){2-5}
\cmidrule(lr){6-9}

\textbf{Repair Scenarios} &
\textbf{\#MF} & \textbf{\#SF} & \textbf{\#SH} & \textbf{\#SL} &
\textbf{\#MF} & \textbf{\#SF} & \textbf{\#SH} & \textbf{\#SL} \\
\midrule

ChatRepair &
-- & 76 & -- & -- &
-- & -- & -- & 48 \\

ThinkRepair &
-- & 98 & 78 & 52 &
-- & 107 & 81 & 47 \\

RepairAgent &
7 & 83 & 71 & 51 &
6 & 68 & 65 & 48 \\

ReinFix\textsubscript{\scriptsize GPT4o} &
22 & 124 & 93 & 57 &
15 & 130 & 103 & 56 \\

\viberepair\textsubscript{\scriptsize GPT4o} &
\textbf{30} & 117 & 80 & 49 &
\textbf{23} & \textbf{144} & 103 & 53 \\

\cellcolor{gray!20}\viberepair\textsubscript{\scriptsize miniR\_GPT4o} &
\cellcolor{gray!20}\textbf{39} & \cellcolor{gray!20}\textbf{135} & \cellcolor{gray!20}\textbf{88} & \cellcolor{gray!20}53 &
\cellcolor{gray!20}\textbf{23} & \cellcolor{gray!20}\textbf{155} & \cellcolor{gray!20}\textbf{108} & \cellcolor{gray!20}55 \\
\bottomrule
\end{tabular}
\end{adjustbox}

\label{tab:repair_scenarios}
\end{table*}

\subsubsection{Repair Scenarios.} 
We further analyze repair effectiveness across different repair scenarios on Defects4J v1.2 and Defects4J v2.0, as summarized in Table \ref{tab:repair_scenarios}. Since ChatRepair and ThinkRepair primarily focus on SF bugs and provide limited support for more complex scenarios, we conduct a more detailed comparison across repair scenarios to better understand how different approaches perform beyond simple single-function repairs.

On Defects4J v1.2, \viberepair\textsubscript{\scriptsize miniR\_GPT4o} consistently achieves the strongest performance across all repair scenarios. In particular, it repairs 39 MF bugs, compared to 22 repaired by ReinFix\textsubscript{\scriptsize GPT4o} and 7 repaired by RepairAgent, demonstrating a clear advantage in handling bugs that span multiple functions. \viberepair\textsubscript{\scriptsize GPT4o} also performs competitively on MF bugs, repairing 30 MF bugs, which exceeds the performance of all baselines except \viberepair\textsubscript{\scriptsize miniR\_GPT4o}. In the SF scenario, \viberepair\textsubscript{\scriptsize miniR\_GPT4o} repairs 135 bugs, outperforming ReinFix (124) and ThinkRepair (98), while \viberepair\textsubscript{\scriptsize GPT4o} repairs 117 SF bugs, remaining comparable to ReinFix and stronger than other baselines. For SH and SL bugs, \viberepair\textsubscript{\scriptsize miniR\_GPT4o} also shows consistent advantages, repairing 88 SH and 53 SL bugs. Although the number of repaired SL bugs is slightly lower than ReinFix (57), \viberepair compensates by repairing substantially more MF and SF bugs, indicating a broader repair capability across different scenarios.

The results on Defects4J v2.0 exhibit similar trends. \viberepair\textsubscript{\scriptsize miniR\_GPT4o} repairs 23 MF bugs, exceeding ReinFix (15) and RepairAgent (6), while \viberepair\textsubscript{\scriptsize GPT4o} repairs the same number of MF bugs, further confirming that the specification-centric approach is effective even without the Reasoning Component. In the SF scenario, \viberepair\textsubscript{\scriptsize miniR\_GPT4o} repairs 155 bugs, outperforming all baselines, and \viberepair\textsubscript{\scriptsize GPT4o} repairs 144 bugs, surpassing ReinFix (130) and ThinkRepair (107). For SH and SL bugs, \viberepair\textsubscript{\scriptsize miniR\_GPT4o} remains competitive, repairing 108 SH and 55 SL bugs. While SL repairs are slightly fewer than ReinFix (56), \viberepair consistently repairs more MF and SF bugs, highlighting its strength in handling more complex and diverse bug scenarios.

\subsubsection{Unique Fixes.}
\begin{figure}[!htp]
    \centering
    \begin{minipage}[t]{0.48\textwidth}
        \centering
        \includegraphics[width=\textwidth]{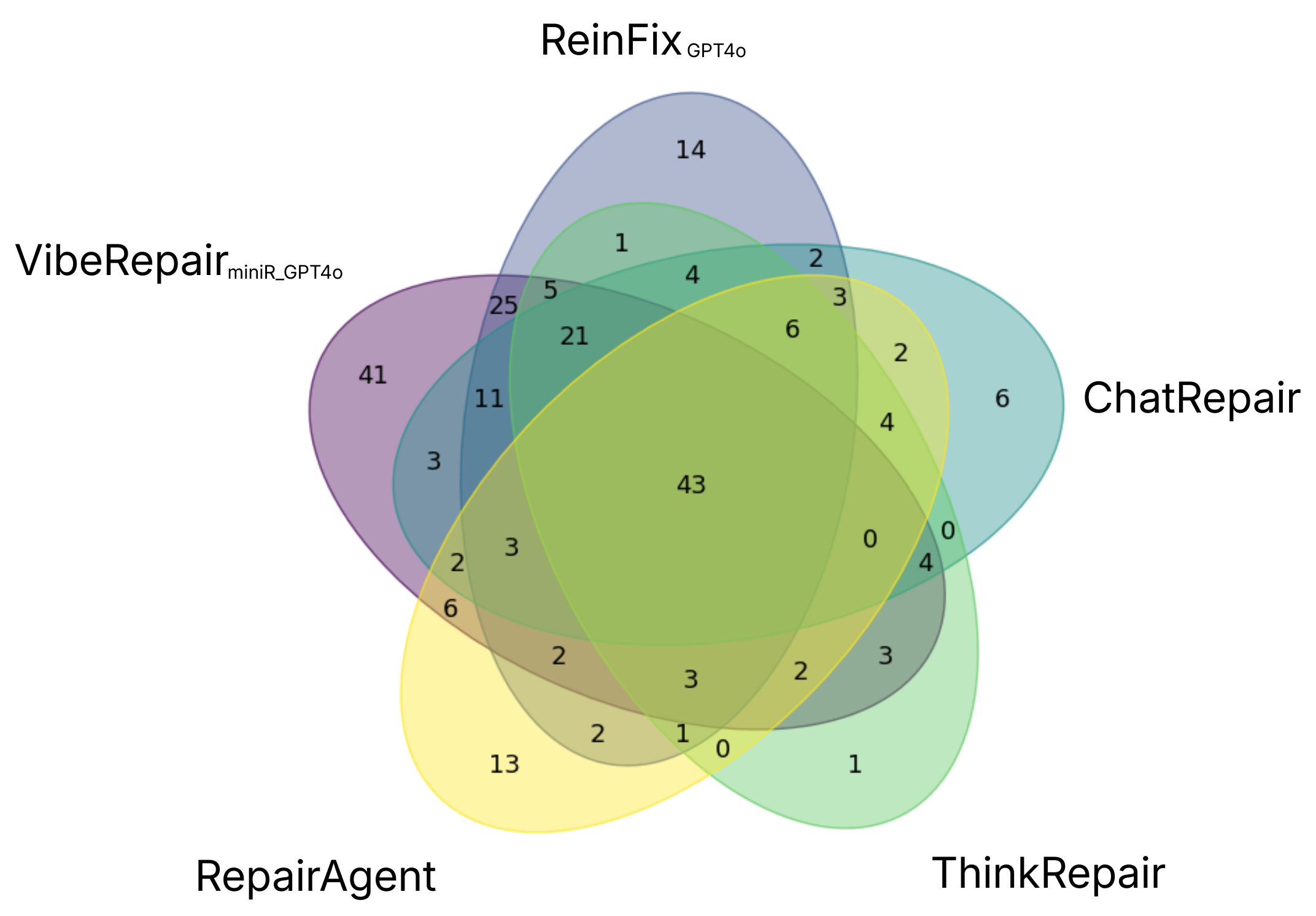}
        \subcaption{Defects4J v1.2}
        \label{fig:venn_d4j1.2}
    \end{minipage}\hfill
    \begin{minipage}[t]{0.48\textwidth}
        \centering
        \includegraphics[width=\textwidth]{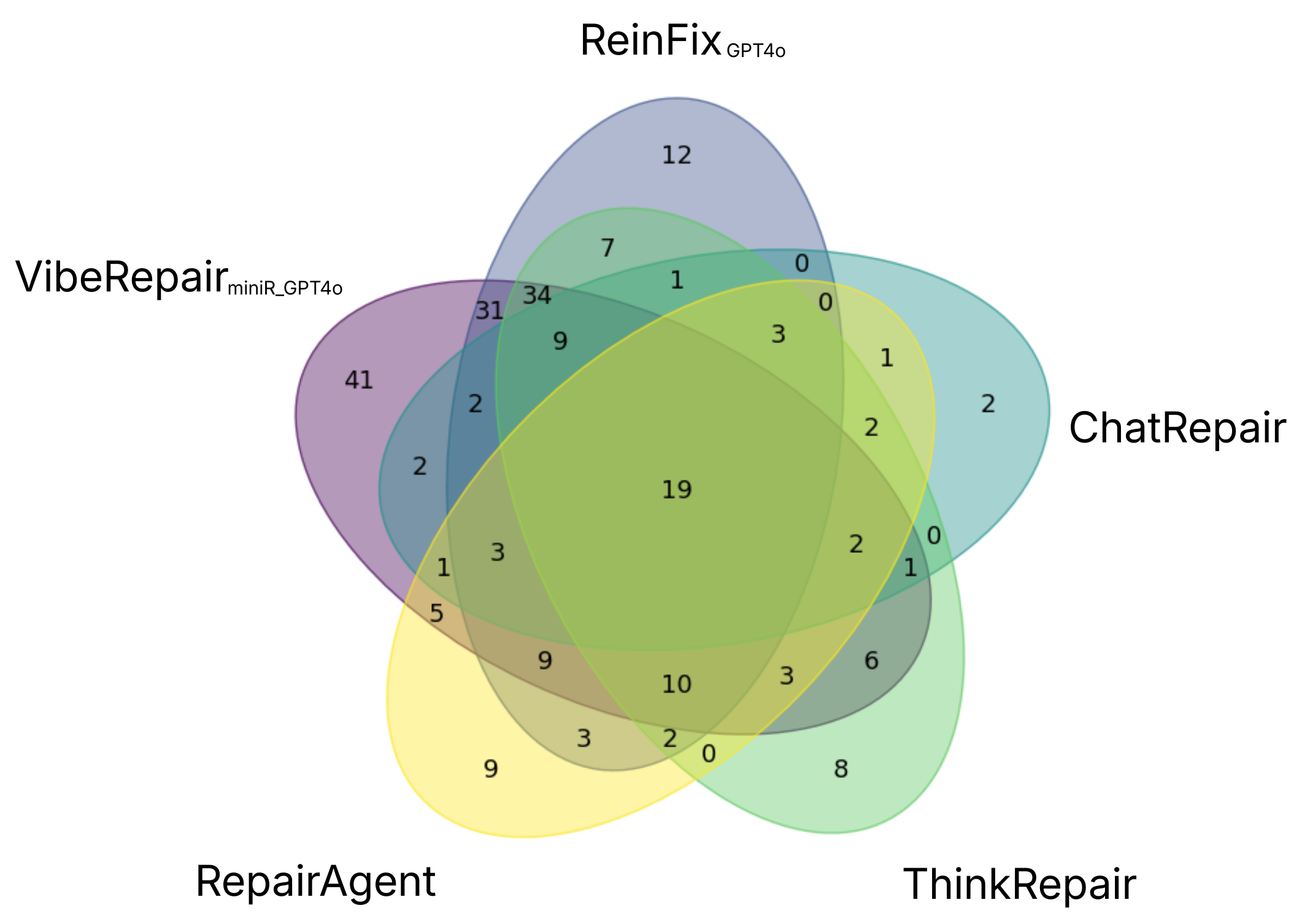}
        \subcaption{Defects4J v2.0}
        \label{fig:venn_d4j2.0}
    \end{minipage}
    \caption{Bug-fix Venn diagrams on Defects4J, showing the overlap of bugs repaired by different tools.}
    \label{fig:venn_d4j}
\end{figure}

We further analyse the unique fixes produced by different APR tools on Defects4J v1.2 and Defects4J v2.0. On Defects4J v1.2 (Fig.~\ref{fig:venn_d4j}\subref{fig:venn_d4j1.2}), \viberepair\textsubscript{\scriptsize miniR\_GPT4o} produces 41 unique fixes, substantially more than any baseline. In comparison, ReinFix\textsubscript{\scriptsize GPT4o} yields 14 unique fixes, i.e., \viberepair repairs 27 more bugs exclusively than ReinFix. The other baselines contribute far fewer unique fixes (ChatRepair: 6; ThinkRepair: 1; RepairAgent: 13). These results indicate that many of the bugs repaired by \viberepair remain unaddressed by existing LLM-based APR tools.

A similar pattern is observed on Defects4J v2.0 (Fig.~\ref{fig:venn_d4j}\subref{fig:venn_d4j2.0}). \viberepair\textsubscript{\scriptsize miniR\_GPT4o} again achieves the highest number of unique fixes, repairing 41 bugs that none of the other tools can fix. By contrast, ReinFix\textsubscript{\scriptsize GPT4o} produces 12 unique fixes, so \viberepair repairs 29 more unique bugs than ReinFix on this dataset. The remaining baselines contribute only a small number of exclusive repairs (ChatRepair: 2; ThinkRepair: 8; RepairAgent: 9).

Overall, this analysis shows that \viberepair’s gains are not explained solely by overlap with existing tools. Instead, \viberepair expands the repair space by fixing many bugs that prior APR techniques do not address, supporting the effectiveness of its specification-centric approach.

\begin{tcolorbox}
\textbf{Answer to RQ1:} \viberepair consistently outperforms existing APR approaches on Defects4J v1.2 and v2.0, repairing more bugs with a much smaller patch space. It also produces a notable number of unique correct fixes and maintains strong performance across different repair scenarios, demonstrating the effectiveness of specification-centric repair.
\end{tcolorbox}

\subsection{RQ2 - Effectiveness–Cost Trade-offs of Reasoning Component Usage Strategies}
To study how the optional Reasoning Component affects repair effectiveness and cost, and to understand how to choose an appropriate usage strategy, we evaluate three \viberepair variants under the same base model on two benchmarks. On Defects4J, we use GPT-4o as the base model and consider: (i) \viberepair\textsubscript{\scriptsize GPT4o} (never enabling the Reasoning Component), (ii) \viberepair\textsubscript{\scriptsize miniR\_GPT4o} (enabling it only when the default repair process fails), and (iii) \viberepair\textsubscript{\scriptsize maxR\_GPT4o} (always enabling it). On RWB v1.0, we use GPT-4 as the base model and evaluate the corresponding three variants: \viberepair\textsubscript{\scriptsize GPT4}, \viberepair\textsubscript{\scriptsize miniR\_GPT4}, and \viberepair\textsubscript{\scriptsize maxR\_GPT4}.

\begin{table*}[h]
\caption{Repair effectiveness, average time (seconds), and cost under different reasoning component usage strategies on Defects4J v2.0 and RWB v1.0.}
\centering
\footnotesize
\renewcommand{\arraystretch}{1.3}
\setlength{\extrarowheight}{2pt}
\setlength{\tabcolsep}{4.5pt}  
\captionsetup{skip=6pt}

\begin{adjustbox}{max width=\textwidth}
\begin{tabular}{p{2.8cm} ccc} 
\toprule
\textbf{Benchmarks} &
\multicolumn{3}{c}{\textbf{Defects4J v2.0}} \\
\cmidrule{2-4}

\textbf{Variants} &
\textbf{Correct Fixes} & \textbf{Avg. Time (s)} & \textbf{Avg. \$}  \\
\midrule

\viberepair\textsubscript{\scriptsize GPT4o} &
167 & 107.3 & 0.066 \\

\viberepair\textsubscript{\scriptsize miniR\_GPT4o} &
178 & 112.8 & 0.074  \\

\viberepair\textsubscript{\scriptsize maxR\_GPT4o} &
174 & 161.0 & 0.221  \\

\midrule

\textbf{Benchmarks} &
\multicolumn{3}{c}{\textbf{RWB v1.0}} \\
\cmidrule{2-4}

\textbf{Variants} &
\textbf{Correct Fixes} & \textbf{Avg. Time (s)} & \textbf{Avg. \$}  \\
\midrule

\viberepair\textsubscript{\scriptsize GPT4} &
10 & 278.8 & 0.491 \\

\viberepair\textsubscript{\scriptsize miniR\_GPT4} &
13 & 303.4 & 0.653  \\

\viberepair\textsubscript{\scriptsize maxR\_GPT4} &
13 & 358.4 & 1.738 \\
\bottomrule
\end{tabular}
\end{adjustbox}

\label{tab:usage_strategies}
\end{table*}

Table \ref{tab:usage_strategies} summarizes the repair effectiveness, average time per bug, and average cost per bug for different strategies. On Defects4J v2.0, \viberepair\textsubscript{\scriptsize GPT4o} repairs 167 bugs with an average time of 107.3 seconds and an average cost of \$0.066. Selectively enabling the Reasoning Component (\viberepair\textsubscript{\scriptsize miniR\_GPT4o}) increases the number of correct fixes to 178, while only slightly increasing the average time (112.8 seconds) and cost (\$0.074). In contrast, always enabling reasoning (\viberepair\textsubscript{\scriptsize maxR\_GPT4o}) repairs 174 bugs, but incurs substantially higher average time (161.0 seconds) and cost (\$0.221). On RWB v1.0, \viberepair\textsubscript{\scriptsize GPT4} repairs 10 bugs, while both \viberepair\textsubscript{\scriptsize miniR\_GPT4} and \viberepair\textsubscript{\scriptsize maxR\_GPT4} repair 13 bugs, with the always-enabled configuration again incurring noticeably higher time and cost.

Taken together, these results indicate that selectively enabling the Reasoning Component only when the default repair process fails provides the most effective and cost-efficient strategy. Across both benchmarks, this strategy consistently achieves the highest or near-highest repair effectiveness while avoiding the substantial overhead introduced by always-on reasoning. Notably, always enabling the Reasoning Component does not further improve repair effectiveness and, in some cases, yields fewer correct fixes than selective enabling. This suggests that excessive auxiliary information can negatively affect the focus of LLM during repair, supporting the need for controlled and targeted use of the Reasoning Component.

Additionally, we observe that RWB v1.0 lacks failing test cases or detailed failure feedback, under which enabling the Reasoning Component leads to further improvements in repair effectiveness. Without such information, inferring the correct intended behavior becomes more difficult, which limits the effectiveness of specification repair based solely on execution feedback. In this setting, enabling the Reasoning Component is particularly beneficial, as it allows \viberepair to leverage additional project-level information to support intent inference.

\begin{tcolorbox}
\textbf{Answer to RQ2:} Enabling the reasoning component only when the default repair process fails achieves the best balance between repair effectiveness and cost. In scenarios where failing test case information is limited or unavailable, activating the reasoning component becomes particularly beneficial, whereas always enabling it incurs higher cost without consistent additional gains.
\end{tcolorbox}

\subsection{RQ3 - Generalizability Study}
To evaluate the generalizability of \viberepair beyond a single LLM, we conduct an additional study using multiple LLMs other than GPT-4o. Following prior studies and to reduce the risk of data leakage, we perform the evaluation on RWB v1.0 using GPT-4 and GPT-3.5, and on RWB v2.0 using DeepSeek-Coder. For fair comparison, all baseline APR tools are evaluated using the same underlying LLMs as \viberepair on each dataset. This setup allows us to assess whether the proposed approach remains effective across different model backbones rather than being tailored to a specific LLM.

\begin{table*}[h]
\caption{Repair results (Correct Fixes) of \viberepair and baseline approaches across different LLMs on RWB benchmarks.}
\centering
\small
\renewcommand{\arraystretch}{1.3}
\setlength{\extrarowheight}{2pt}
\setlength{\tabcolsep}{4.5pt}  
\captionsetup{skip=6pt}

\begin{adjustbox}{max width=\textwidth}
\begin{tabular}{p{2cm} ccc ccc} 
\toprule
\textbf{Benchmarks} &
\multicolumn{3}{c}{\textbf{RWB v1.0}} &
\multicolumn{3}{c}{\textbf{RWB v2.0}} \\
\cmidrule(lr){2-4}
\cmidrule(lr){5-7}

\textbf{APR Tool} &
\textbf{\viberepair\textsubscript{\scriptsize miniR\_GPT4}} & \textbf{ReinFix\textsubscript{\scriptsize GPT4}} & \textbf{\viberepair\textsubscript{\scriptsize miniR\_GPT3.5}} & \textbf{\viberepair\textsubscript{\scriptsize miniR\_DSC}} &
\textbf{ReinFix\textsubscript{\scriptsize DSC}} & \textbf{ThinkRepair\textsubscript{\scriptsize DSC}} \\
\textbf{LLM} &
\textbf{GPT4} & \textbf{GPT4} & \textbf{GPT3.5} & \textbf{DeepSeek} &
\textbf{DeepSeek} & \textbf{DeepSeek} \\
\midrule

Cli &
2 & 2 & 2 & 2 & 2 & 2 \\

Codec &
2 & 2 & 2 & 1 & 2 & 1 \\

Compress &
1 & 1 & 1  &  3 & 1 &  -  \\

Jsoup &
4 & 2 & 2  &  2 & 2 & 2 \\

Lang &
4 & 3 & 2  & 2 & 3 & 3 \\
\midrule

\textbf{\#Total}  &
13 & 10 & 9 & 10 & 10  & 8  \\
\bottomrule
\end{tabular}
\end{adjustbox}

\label{tab:generalizability}
\end{table*}

Table \ref{tab:generalizability} summarizes the repair results across different models and datasets. On RWB v1.0, when using GPT-4, \viberepair\textsubscript{\scriptsize miniR\_GPT4} repairs 13 bugs, compared to 10 bugs repaired by ReinFix\textsubscript{\scriptsize GPT4}. When using the weaker GPT-3.5 model, \viberepair\textsubscript{\scriptsize miniR\_GPT3.5} repairs 9 bugs, indicating that the proposed approach remains effective even with a less capable LLM, although the absolute repair numbers decrease.

On RWB v2.0, evaluated with DeepSeek-Coder, \viberepair\textsubscript{\scriptsize miniR\_DSC} repairs 10 bugs, matching the performance of ReinFix\textsubscript{\scriptsize DSC} and outperforming ThinkRepair\textsubscript{\scriptsize DSC}, which repairs 8 bugs. Although the repair effectiveness is the same, \viberepair achieves this result with a maximum patch space size of 5 × 3, whereas existing approaches rely on substantially larger patch exploration budgets.

\begin{tcolorbox}
\textbf{Answer to RQ3:} \viberepair generalizes well across different LLM backbones, including GPT-4, GPT-3.5, and DeepSeek-Coder, achieving competitive or superior repair effectiveness compared to state-of-the-art baselines.
\end{tcolorbox}

\section{THREATS TO VALIDITY}
\textbf{Internal Validity.} A potential threat to internal validity is data leakage from LLM training data. To mitigate this risk, we follow prior work and evaluate \viberepair on bug cases that were collected after the known cutoff dates of the LLM training corpora. Another threat arises from the inherent nondeterminism of LLM-based approaches, which may produce different patches across runs. To ensure a fair comparison with existing methods, we strictly control the experimental setup when comparing against baselines. In particular, \viberepair uses the same base LLM version as ReinFix, adopts identical fault localization assumptions, and applies the same localization results. When the optional Reasoning Component is enabled, the example dataset used to provide repair guidance is identical to that used in ReinFix, and this dataset has no overlap with any evaluation benchmarks. When evaluating the generalizability across different LLMs, we further control experimental variables by using the same prompts, repair settings, and evaluation procedures across all models. This design ensures that observed performance differences are attributable to model capabilities rather than inconsistencies in configuration or prompt design. Finally, determining correct patches requires manual inspection, which may introduce subjectivity. To mitigate this threat, we follow the criteria and guidelines established in prior APR studies and compare the generated patches against the correct patches reported in those studies.

\textbf{External Validity.} A potential threat to external validity is whether the observed results generalize across a broader range of programs and bug types. While \viberepair may not cover all possible repair scenarios, we follow common practice and conduct extensive evaluations on Defects4J and RWB, two widely used benchmarks that span multiple projects, bug patterns, and repair scenarios. Another external validity concern relates to cost estimation. The reported repair cost is calculated based on token usage and publicly available pricing information at the time of evaluation. In practice, deployment environments, pricing schemes, and model availability may change over time, potentially affecting absolute cost figures. Nevertheless, our comparisons focus on relative cost trends across different usage strategies under the same pricing assumptions, which remain informative for understanding the trade-offs introduced by the proposed approach.
\section{RELATED WORK}
APR has been extensively studied as a means to automatically generate patches based on fault localization results, with the goal of reducing the time and effort required to diagnose and fix software defects \cite{le2016history}. Early research primarily focused on traditional APR techniques, which can be broadly categorized into heuristic-based \cite{jiang2018shaping,le2011genprog,martinez2016astor}, constraint-based \cite{durieux2016dynamoth,martinez2018ultra,xiong2017precise}, and template-based approaches \cite{zhang2023gamma,zhu2023tare,jiang2023knod}. While these methods have shown effectiveness on specific bug patterns, they often suffer from limited generality, strong dependence on manually designed rules or templates, and difficulty in handling complex or previously unseen defects. To address these limitations, learning-based APR techniques were proposed, many approaches formulate patch generation as a neural machine translation task, learning latent bug-fixing patterns from large-scale code repositories \cite{chen2019sequencer,jiang2021cure,li2020dlfix}. Although these techniques achieve promising results, they rely heavily on historical bug-fixing data, and their performance can be negatively affected by data sparsity and noise in training corpora.

With the rapid progress of LLMs, which are pre-trained on massive collections of open-source code, recent research has increasingly explored LLM-based APR. Early work on LLM-based APR primarily focused on direct code generation. AlphaRepair \cite{xia2022less} introduced an infilling-style repair paradigm that masks buggy lines and uses LLMs to complete the code based on surrounding context, demonstrating clear advantages over earlier NMT-based approaches. Similarly, Prenner et al. \cite{prenner2022can} and Kolak et al. \cite{kolak2022patch} explored using Codex to generate repaired functions or complete buggy lines. More recently, Xia et al. \cite{xia2023automated} conducted a systematic empirical study of LLM-based APR across different models and prompting strategies.

At the specification level, SpecRover \cite{ruan2024specrover} proposes to extract intended program behavior from source code and use it to guide LLM-based repair, showing that intent-aware reasoning can improve repair outcomes. However, SpecRover relies on issue statements as additional input to infer intended behavior and still performs repair directly at the code level. As a result, its applicability depends on the availability and quality of external natural-language descriptions, which are not always present in practical repair settings.

In contrast to existing approaches that directly operate on buggy code, we propose a specification-centric APR approach that fundamentally shifts the repair process away from direct code manipulation. \viberepair transforms buggy code and failing test cases into a test-based flawed behavior specification, without requiring issue statements or other external documentation. Through iterative interaction with validation failure feedback, the LLM accurately captures the intended behavior and refines this behavior specification. The final patch is then generated as a direct consequence of the repaired specification, distinguishing our approach from prior code-centric and specification-assisted repair methods.

\section{CONCLUSION}
In this paper, we presented \viberepair, a novel specification-centric approach for LLM-based APR. Unlike existing code-centric methods that directly generate patches from buggy code, \viberepair changes the repair process to operate at the level of behavior specifications, explicitly inferring intended behavior and repairing behavior specifications before regenerating code. By transforming buggy programs into flawed behavior specifications, inferring intended behavior and refining behavior specifications through iterative interaction with validation feedback, and regenerating code from repaired specifications, \viberepair ultimately outputs a patch for the buggy code that realigns the program implementation with the developer’s intent.

Our extensive experiments on Defects4J and real-world benchmarks, including comparisons with state-of-the-art baselines, ablation studies, and generalizability evaluations across multiple LLMs, demonstrate that \viberepair consistently outperforms or matches existing APR approaches while requiring a much smaller patch exploration space. These results highlight the effectiveness of repairing programs by explicitly inferring and refining behavioral intent at the specification level, suggesting a promising direction for LLM-based APR.

\section{DATA AVAILABILITY}
The data and tools used to generate the empirical evidence supporting the main claims of this paper are publicly accessible. Defects4J is available from its official repository, while the RWB datasets released by Yin et al. \cite{yin2024thinkrepair} are accessible at \url{https://github.com/vinci-grape/ThinkRepair/}. Our implementation of \viberepair, the experimental scripts, and the repair results are publicly available at \url{https://anonymous.4open.science/r/VibeRepair-9D73/} to support reproducibility.

\bibliographystyle{ACM-Reference-Format}
\bibliography{all}

\end{document}